\documentclass[10pt]{article}

\usepackage{times}
\usepackage{epsfig}
\usepackage{graphicx}
\usepackage{amsmath}
\usepackage{amssymb}
\usepackage{xcolor}
\usepackage[breaklinks=true,bookmarks=false]{hyperref}
\usepackage[margin=1.1in]{geometry} 
\usepackage{caption}
\usepackage{titlesec}
\usepackage{abstract}

\usepackage[breaklinks=true,bookmarks=false]{hyperref}
\emergencystretch=\maxdimen
\begin{document}

\title{\Large wmh\_seg: Transformer based U-Net for Robust and Automatic White Matter Hyperintensity Segmentation across 1.5T, 3T and 7T}

\author{
Jinghang Li\textsuperscript{1} \\ 
jinghang.li@pitt.edu \and
Tales Santini\textsuperscript{1} \\ 
santini.tales@pitt.edu \and
Yuanzhe Huang\textsuperscript{1} \\ 
yuh94@pitt.edu\and
Joseph M. Mettenburg\textsuperscript{2} \\
mettenburgjm@upmc.edu\and
Tamer S. Ibrahim\textsuperscript{1} \\
tibrahim@pitt.edu\and
Howard J. Aizenstein\textsuperscript{1,3} \\
aizen@pitt.edu\and
Minjie Wu\textsuperscript{1,3} \\
miw75@pitt.edu\and
}
\date{} 

\maketitle

\begin{flushleft}
\textsuperscript{1}Department of Bioengineering, University of Pittsburgh, Pittsburgh, Pennsylvania, United States of America  \\
\textsuperscript{2}Department of Radiology, University of Pittsburgh School of Medicine, Pennsylvania, United States of America \\
\textsuperscript{3}Department of Psychiatry, University of Pittsburgh School of Medicine, Pittsburgh, Pennsylvania, United States of America 
\end{flushleft}

\maketitle

\begin{abstract}
White matter hyperintensity (WMH) remains the top imaging biomarker for neurodegenerative diseases. Robust and accurate segmentation of WMH holds paramount significance for neuro imaging studies. The growing shift from 3T to 7T MRI necessitates robust tools for harmonized segmentation across field strengths and artifacts. Recent deep learning models exhibit promise in WMH segmentation but still face challenges, including diverse training data representation and limited analysis of MRI artifacts' impact. To address these, we introduce wmh\_seg, a novel deep learning model leveraging a transformer-based encoder from SegFormer. wmh\_seg is trained on an unmatched dataset, including 1.5T, 3T, and 7T FLAIR images from various sources, alongside with artificially added MR artifacts. Our approach bridges gaps in training diversity and artifact analysis. Our model demonstrated stable performance across magnetic field strengths, scanner manufacturers, and common MR imaging artifacts. Despite the unique inhomogeneity artifacts on ultra-high field MR images, our model still offers robust and stable segmentation on 7T FLAIR images. Our model to date is the first that offers quality white matter lesion segmentation on 7T FLAIR images.\end{abstract}

\section{Introduction}

White matter hyperintensities (WMH) are hyperintense lesion clusters on T2 weighted (T2w) magnetic resonance images (MRI). These hyperintense clusters are commonly observed in the brains of older individuals and those with various neurological disorders such as Alzheimer's disease (AD) \cite{lee2016white} and multiple sclerosis (MS) \cite{polman2011diagnostic}. Although WMH occurrence is common within normal aging, significant lesion volumes can negatively impact the individual’s cognitive functions \cite{inzitari2009changes} \cite{lampe2019lesion}. The presence of WMH is associated with greater likelihood and early presence of clinical dementia, poorer cognitive performance, and more severe clinical symptoms in AD patients  \cite{esiri1999cerebrovascular}\cite{heyman1998cerebral}\cite{schneider2004cerebral}\cite{snowdon1997brain}.   For MS patients, lesions at the critical brain regions can impose unpleasant disease symptoms such as gait disturbances, muscle weakness, and bladder problems \cite{ben2011therapeutics}. As a result, accurate segmentation of WMHs is crucial for brain aging assessment, neurological disease diagnosis, and disease progression monitoring.  

Traditionally, WMH segmentation has been relied on manual efforts that are not only time-consuming, but also subjective, introducing undesired reader variability \cite{prins2004measuring}. To achieve better segmentation of the white matter hyperintensities, semi-automated and fully automatic pipelines were proposed since the early 2000s \cite{admiraal2005fully}\cite{iorio2013white}\cite{schmidt2017lst}\cite{wu2006fully}. However, these pipelines often involve extensive preprocessing steps and fine-tuning of hyper-parameters tailored to specific datasets, thus making them less suitable for large multi-site studies. Further, distinctive 7T inhomogeneity artifacts (insufficient signal in cerebellum and temporal lobe, etc.) \cite{zwanenburg2010fluid} impose additional challenges for these traditional segmentation pipelines. As brain imaging continuously advances from 3T to 7T MR image acquisition, offering improved resolution \cite{kang2009hypertension}, enhanced soft tissue contrast \cite{chen2023paired}, and increased statistical strength \cite{torrisi2018statistical}, the demand for a robust segmentation tool that adapts well across domains becomes increasingly pressing. Such tool should demonstrate strong generalization across various magnetic field strengths, scanner manufacturers, and commonly encountered MRI artifacts. Such segmentation model would prove especially valuable for studies incorporating both 3T and 7T images, offering harmonized segmentation outputs across different imaging scenarios.

In recent years, various deep learning models have shown great promise for segmenting WMHs \cite{ronneberger2015u}\cite{wu2019skip}. These deep learning-based segmentation models have strong capacity to capture the inductive bias within the heterogenous training images. Despite the success of deep learning-based methods for WMH segmentation, there are still several challenges that need to be addressed. One of the major challenges is the training dataset diversity. Although, one publicly available WMH segmentation dataset contains MR images acquired at 1.5 Tesla and 3 Tesla \cite{kuijf2019standardized}. They did not include 7 Tesla MR images. Furthermore, few published works have addressed the impact of commonly seen MRI artifacts, such as noise, and bias field inhomogeneity, on the performance of segmentation models. As a result, it remains essential to develop a deep learning model that is trained on more diverse datasets that can better capture the variability in imaging conditions and to investigate the impact of these artifacts on segmentation performance.

To combat the above-mentioned issues, we implemented wmh\_seg, a deep learning model that leverages the transformer-based encoder from recently published state-of-the-art SegFormer \cite{xie2021segformer}. wmh\_seg was trained on an unparalleled training dataset that includes Fluid Attenuated Inverse Recovery (FLAIR) images acquired at 1.5 Tesla, 3 Tesla, and 7 Tesla magnetic field strengths across 4 institutes and 3 MR scanner manufacturers. Further, we implemented various data augmentations to improve the robustness of wmh\_seg for WMH segmentation. Our work, to date, is one of a kind that incorporates training images from the most diverse sources. In addition, to the best of our knowledge, we are the first group to include 7T FLAIR training images for WMH segmentation. 

\section{Related Work}
\subsection{Classical White Matter Lesion Segmentation}

FreeSurfer Image Analysis Suite \cite{fischl2012freesurfer} is a widely acknowledged neuro image segmentation tool. It offers extensive parcellation for the cortical and subcortical brain structures. Although it is not optimized for segmenting white matter lesions, it can detect the abnormal hypointensities on T1 weighted (T1w) MR images by leveraging the built-in T1w probabilistic tissue map and image intensity information. It is important to note that the T1w MR images do not offer the best contrast for white matter lesions and it is common practice to perform white matter lesion segmentation on T2w FLAIR images for better quantification. 

To combat the low lesion contrast issue on T1w images, FSL (FMRIB Software Library) and SPM (Statistical Parametric Mapping) developed multi-modal white matter lesion segmentation toolboxes \cite{griffanti2016bianca}\cite{schmidt2017lst}. These tools, however, are not fully automated and often require user inputs and hyper-parameter adjustment, which creates significant barrier for researchers who are not familiar with the technical details. 

\subsection{Deep Learning Approaches}
The U-net model proposed in 2015 is a significant milestone for image segmentation in the medical domain \cite{ronneberger2015u}. It is a convolutional neuro network (CNN) that used hierarchical encoder and decoder as well as skip connections to capture the image features at both global and local scales. Since its publication, many U-net model variants have been applied to various medical image tasks such as brain, vessel, and lung segmentation, lesion detection and classification. 

These deep learning models can learn the inductive bias from the training images and offer an end-to-end model that requires no user inputs and hyperparameter tuning. In 2017, a white matter lesion segmentation challenge was held to promote the development of the most robust method for white matter lesion detection \cite{kuijf2019standardized}. The challenge winning team leveraged a CNN U-Net model and showed high quality segmentation result across scanner manufacturers, image resolution, and demonstrated stable segmentations on MR images acquired at 1.5T and 3T respectively \cite{park2021white}. 

\subsection{Contribution}
wmh\_seg builds upon a solid foundation on prior works that explored the most efficient network architectures for lesion segmentation. Our goal is to offer a relatively robust WMH segmentation toolbox that generalizes well across a wide range of data source. Our model will be the first that’s being trained on 7T FLAIR images for white matter hyperintensity segmentation. In our comparison, our model outputs stable segmentation across MR image artifacts. Specifically, it offers accurate WMH quantification on 7T FLAIR images. The trained model will be made publicly available after the manuscript is accepted. The segmentation project page can be found at: https://github.com/jinghangli98/wmh\_seg 

\section{Material and Methods}

\subsection{Data Distribution}

\begin{figure}[h]
\begin{center}
   \includegraphics[width=1\linewidth]{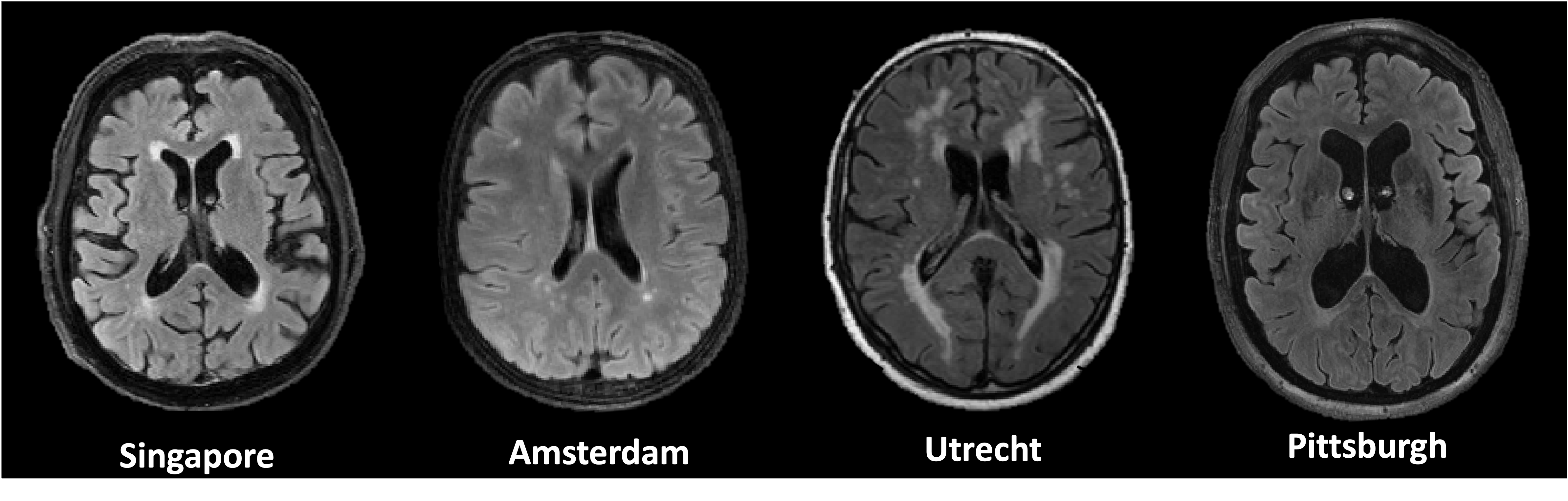}
\end{center}
   \caption{Axial slices of T2 weighted FLAIR images acquired different institutes. These axial slices illustrate the significant MR image variability due to the differences in MR scanner manufacturer, magnetic field strength, and sequence acquisition protocol.}
\label{fig:fig1}
\end{figure}

To further improve the diversity upon the publicly available WMH segmentation dataset, we incorporated additional 100 T2 weighted 7 Tesla normal aging FLAIR images acquired at the University of Pittsburgh. In this work, we had 270 unique FLAIR scans from many different universities and institutes, including University of Pittsburgh, UMC Utrecht, NUHS Singapore, and VU Amsterdam, to train and evaluate our transformer based U-net model for WMH segmentation \cite{kuijf2019standardized}. Figure \ref{fig:fig1} showcases the acquired T2w FLAIR axial image differences at the 4 sites. Table 1 below provides the detailed MR information of the dataset \cite{kuijf2019standardized}. All 7T FLAIR WMH masks were manually created using ITK-Snap, each of which was carefully reviewed by our trained radiologist. 
\begin{table}[h!]
\centering
\caption{Summary of FLAIR Image Resolutions Across Different Institutes}
\label{tab:scanner_specs}
\begin{tabular}{llcc}
\hline
Institute & Scanner & \#scans & Resolution \\
\hline
University of Pittsburgh & 7 T Siemens Magnetom sTx & 100 & 0.75×0.75×1.5 mm\textsuperscript{3} \\
UMC Utrecht & 3 T Philips Achieva & 50 & 0.96×0.95×3.00 mm\textsuperscript{3} \\
NUHS Singapore & 3 T Siemens TrioTim & 50 & 1.0×1.0×3.0 mm\textsuperscript{3} \\
VU Amsterdam & 3 T GE Signa HDxt & 50 & 0.98×0.98×1.2 mm\textsuperscript{3} \\
 & 3 T Philips Ingenuity & 10 & 1.04×1.04×0.56 mm\textsuperscript{3} \\
 & 1.5 T GE Signa HDxt & 10 & 1.21×1.21×1.30 mm\textsuperscript{3} \\
\hline
\end{tabular}
\end{table}

This study analyzes collected data from current on-going neuroimaging studies as well as publicly available open access data. The 7T imaging studies were approved by IRBs at the University of Pittsburgh. All subjects were given written informed consent. No ethical approval was required for de-identified open access data.

\subsection{Data Augmentation}

\begin{figure}[h]
\begin{center}
   \includegraphics[width=1\linewidth]{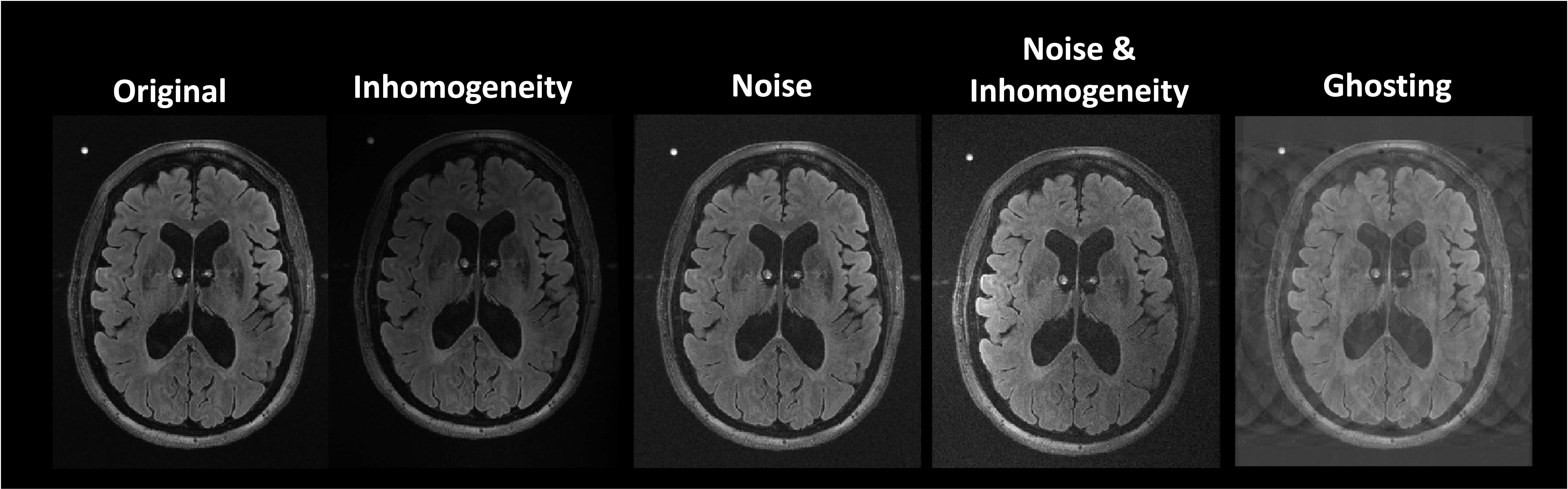}
\end{center}
   \caption{Commonly observed MRI artifacts implemented for data augmentation using torchio \cite{perez2021torchio}}
\label{fig:fig2}
\end{figure}

Properly applied data augmentation has demonstrated its ability to bolster the robustness and elevate the performance of deep learning models \cite{garcea2023data}\cite{taylor2018improving}. To extend the generalizability of our segmentation model beyond “quality” images, we implemented common MRI artifacts on our training T2w FLAIR images via TorchIO \cite{perez2021torchio}. These MR artifacts include noise, inhomogeneity, and ghosting (Figure \ref{fig:fig2})

In addition to the individual artifact, we also combined noise with inhomogeneity artifact to simulate the scenario where both the transmit and the receive MR head coil are not in the optimal condition. Each unique FLAIR scan is accompanied by 4 “corrupted” images; in total we have 1350 FLAIR images for training and testing. The training and testing datasets are randomly split with an 80-20 ratio. We evaluated the performance of the model using the Dice score, which measures the overlap between the predicted masks and ground truth WMH masks. Furthermore, we tested the model performance on 1.5T, 3T, and 7T FLAIR images and compared the segmentation results with the WMH segmentation challenge winning team submission model. 

\subsection{Network Architecture}

We leveraged the encoder component proposed in the SegFormer model \cite{xie2021segformer} to build our wmh\_seg model. Unlike traditional CNNs, wmh\_seg’s encoder employs an attention mechanism that significantly improves the receptive fields of the deep learning model \cite{raghu2021vision}. The encoder first creates the overlap patch embedding, which is then fed into a subsequent transformer block. To reduce the computational complexity of the self-attention mechanism, the implemented encoder block reduced the dimensions of the query and key vectors in the efficient attention layer by a factor of R, resulting in a decreased complexity from $O(N^2)$ to  $O(\frac{N^2}{R})$, where N is the length of the flattened tensor (1 x (H x W) x C) from the patch embedding block (see Figure \ref{fig:fig3}). Instead of implementing a positional encoding vector, a convolutional feed-forward block was used following the efficient self-attention output to encode the positional information. This approach can be resolution-agnostic and was shown to be effective during model inference \cite{xie2021segformer}.

To improve the model complexity, instead of using multi-layer perceptron for the decoder block proposed in the original SegFormer model, we opted hierarchical convolutional layers to upsample the encoded features and generated the segmentation mask. 

\subsection{Implementation}
\begin{figure}[h]
\begin{center}
   \includegraphics[width=1\linewidth]{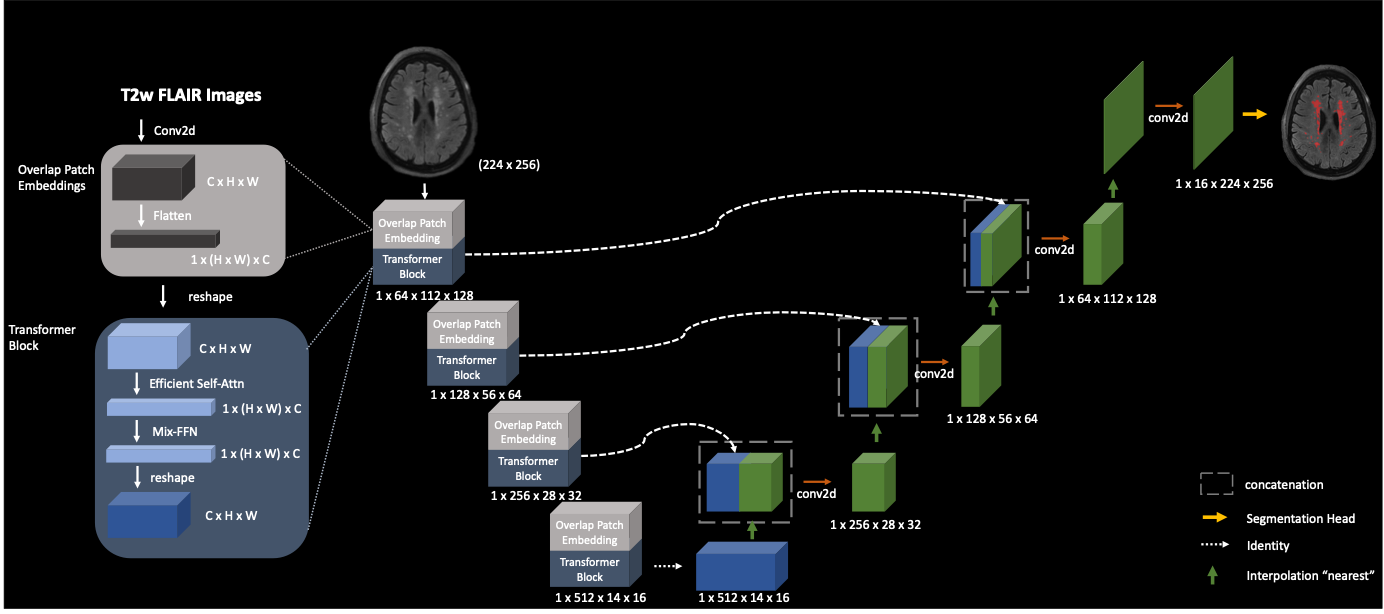}
\end{center}
   \caption{wmh\_seg model architecture. The model consists of hierarchical transformer encoder and convolutional decoder.}
\label{fig:fig3}
\end{figure}

The implemented wmh\_seg largely resembles the U-net architecture (Figure \ref{fig:fig3}). In the encoder component, we leveraged the transformer blocks proposed in \cite{xie2021segformer}. In the decoder component, we utilized convolutional layers to up-sample the encoded image features. The detailed feature dimensions and channel numbers are illustrated in Figure \ref{fig:fig3}. 
To achieve rather robust segmentation results across diverse image sources, we adopted the following loss function during training:
\begin{equation}
    \mathcal{L} = \mathcal{L}_{\text{BCE}} + \mathcal{L}_{\text{Dice}}
\end{equation}
\begin{equation}
    \mathcal{L}_{\text{BCE}} = -\frac{1}{N} \sum_{i=0}^{N} \left( y_i \odot \log(\hat{y}_i) + (1 - y_i) \odot \log(1 - \hat{y}_i) \right)
\end{equation}
\begin{equation}
    \mathcal{L}_{\text{Dice}} = \frac{2 | y_i \odot \hat{y}_i |}{|y_i| \oplus |\hat{y}_i|}
\end{equation}
with $\odot$ being the element-wise multiplication and $\oplus$ being the element-wise addition \cite{rajput4065778robustness}.

We trained our model using the Adam optimizer \cite{kingma2014adam}, with a patience of 2 reduce-on-plateau scheduler for 100 epochs. The initial learning rate was set to $10^{-4}$, and the training batch size was 32. To ensure the image size consistency, we cropped or zero-padded the training images to a size of (256 x 256) and normalized their intensity values between 0 and 1. We implemented wmh\_seg in Python and used Pytorch to facilitate the implementation. All model trainings were conducted on NVDIA A100 GPUs at the University of Pittsburgh Center for Research Computing. The total training time was 46 hours.

\newpage
\section{Results}
We evaluated our model's segmentation performance on diverse T2w FLAIR images, encompassing various magnetic field strengths and artificially introduced artifacts. Additionally, we compared our model outputs with the WMH segmentation challenge winning team's model: pgs, both quantitatively and qualitatively, specifically focusing on lower field strength FLAIR images (1.5T and 3T). We refrained from a quantitative comparison of our model's performance with pgs on 7T FLAIR images due to its lack of training on the higher field strength FLAIR images. Instead, we solely performed a qualitative comparison. Regarding FreeSurfer, we also refrained from quantitative comparison due to its segmentation limitation on the T1w images only, and solely performed a qualitative comparison with FreeSurfer's segmentation on the corresponding T1w images.

\begin{figure}[!htb]
\begin{center}
   \includegraphics[width=1\linewidth]{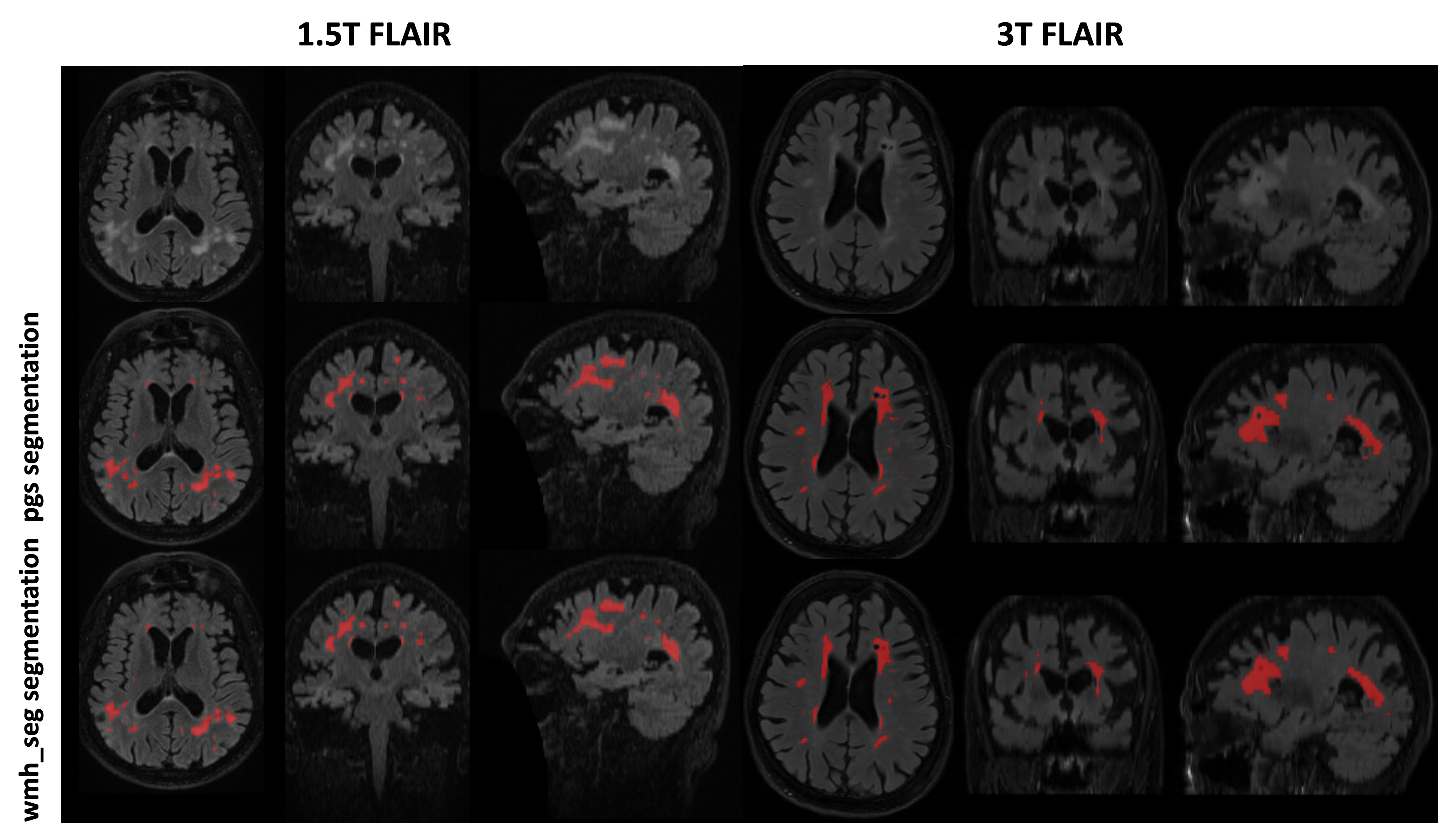}
\end{center}
   \caption{wmh\_seg and pgs segmentation results on 1.5 Tesla, 3 Tesla FLAIR images.}
\label{fig:fig4}
\end{figure}

\begin{figure}[!htb]
\begin{center}
   \includegraphics[width=1\linewidth]{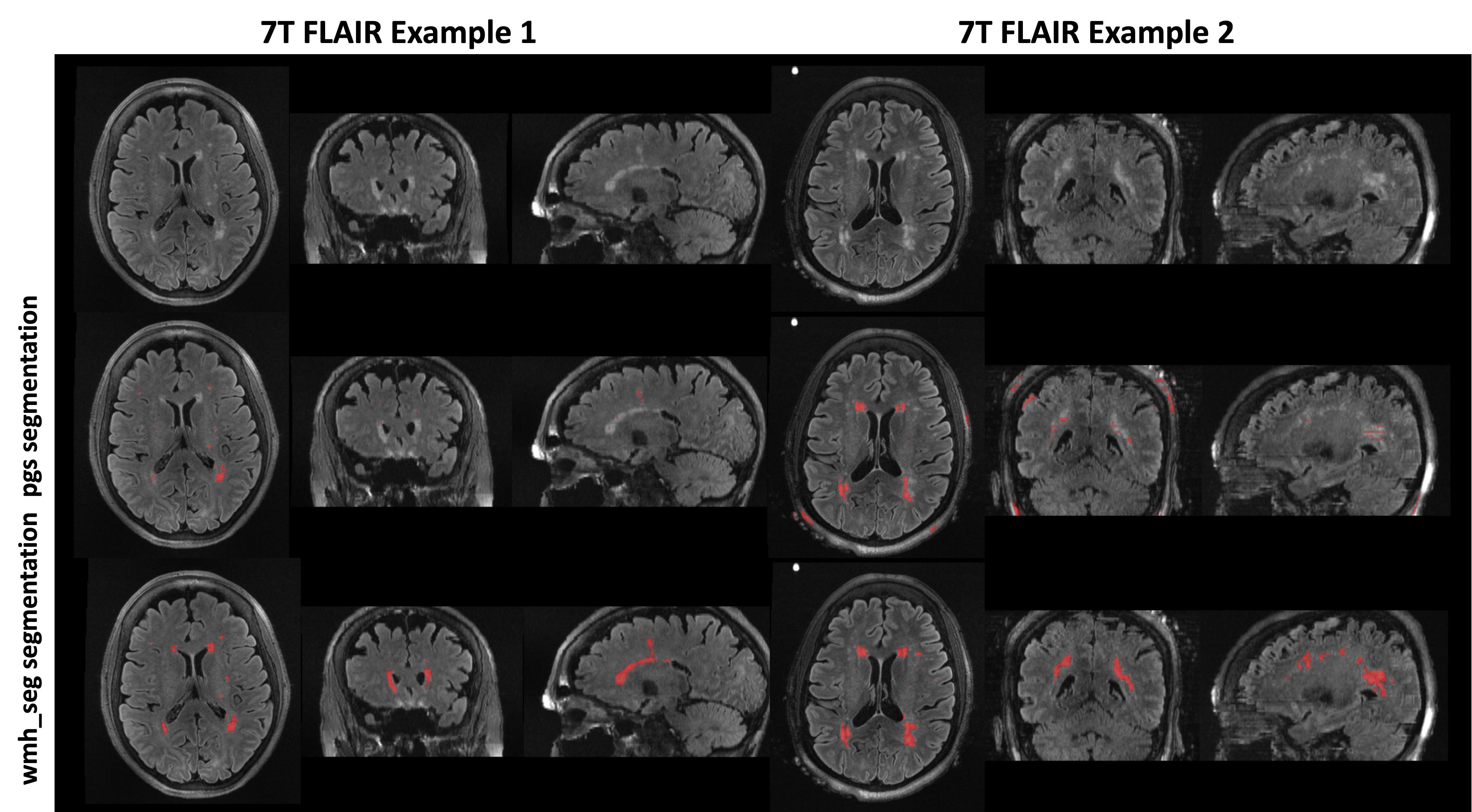}
\end{center}
   \caption{wmh\_seg and pgs segmentation results on 7 Tesla FLAIR images.}
\label{fig:fig5}
\end{figure}

To examine the model's generalizability across different magnetic field strengths, we conducted model inference on randomly selected FLAIR images acquired at 1.5 Tesla (N = 3), 3 Tesla (N = 30), and 7 Tesla (N = 24) from the testing dataset. We also validated the model's performance on FLAIR images with added artifacts. The results shown in figures \ref{fig:fig4} and \ref{fig:fig5} illustrate that the wmh\_seg achieves comparable segmentation quality to the challenge-winning team's model for 1.5T and 3T FLAIR images, while notably outperforming it on the segmentation of 7T FLAIR images. 

\begin{figure}[!htb]
\begin{center}
   \includegraphics[width=1\linewidth]{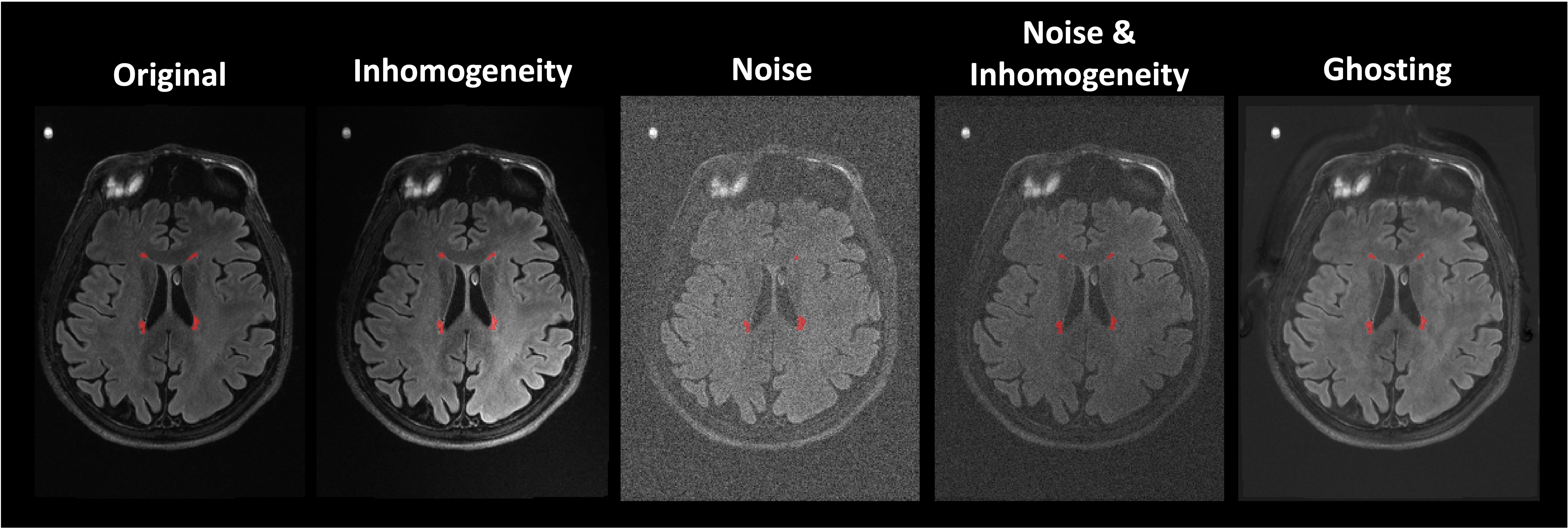}
\end{center}
   \caption{wmh\_seg segmentation on the commonly seen MR artifacts. The segmentation masks largely remain intact and are not impacted by the artificial corruptions.}
\label{fig:fig6}
\end{figure}

\begin{figure}[!htb]
\begin{center}
   \includegraphics[width=1\linewidth]{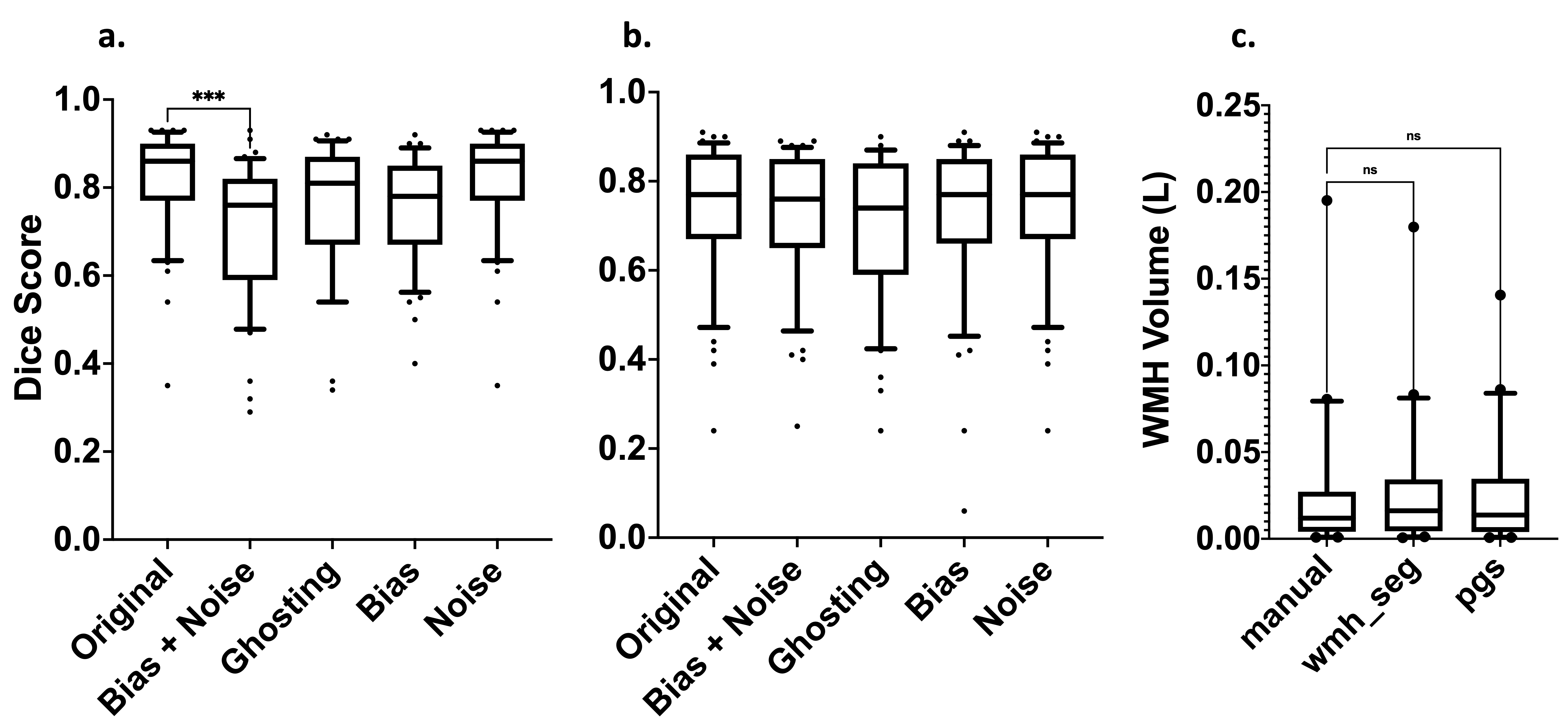}
\end{center}
   \caption{Quantitative segmentation comparison between pgs and wmh\_seg on 1.5T, 3T FLAIR images and images with corresponding artifacts. a) pgs model inference on 1.5T, and 3T FLAIR images. With added artifacts, the segmentation results showed differences on images with bias + noise artifacts (***, $p < 0.0015$). b) wmh\_seg model inference on 1.5T, 3T FLAIR images. c) volumetric comparison shows no differences between manual labeled mask and segmentation outputs on the non-corrupted 1.5T and 3T FLAIR images from the two models.}
\label{fig:fig7}
\end{figure}

Figures \ref{fig:fig6}, and \ref{fig:fig7} show that wmh\_seg is more robust than pgs to common MR artifacts such as inhomogeneity and noise. 

\newpage

\newpage
Our evaluations suggest that incorporating artificial MRI artifacts from diverse sources in the training data can improve the robustness of white matter lesion segmentation. Although, pgs has better dice score evaluation on 1.5T and 3T FLAIR images, our result in figure \ref{fig:fig7} shows that there’s no volumetric differences between the two. However, for 7T FLAIR images, where the inhomogeneity is significantly more prominent than that of 1.5T and 3T images, pgs suffers catastrophic failure (figure \ref{fig:fig5}). 

\begin{figure}[!htb]
\begin{center}
   \includegraphics[width=0.7\linewidth]{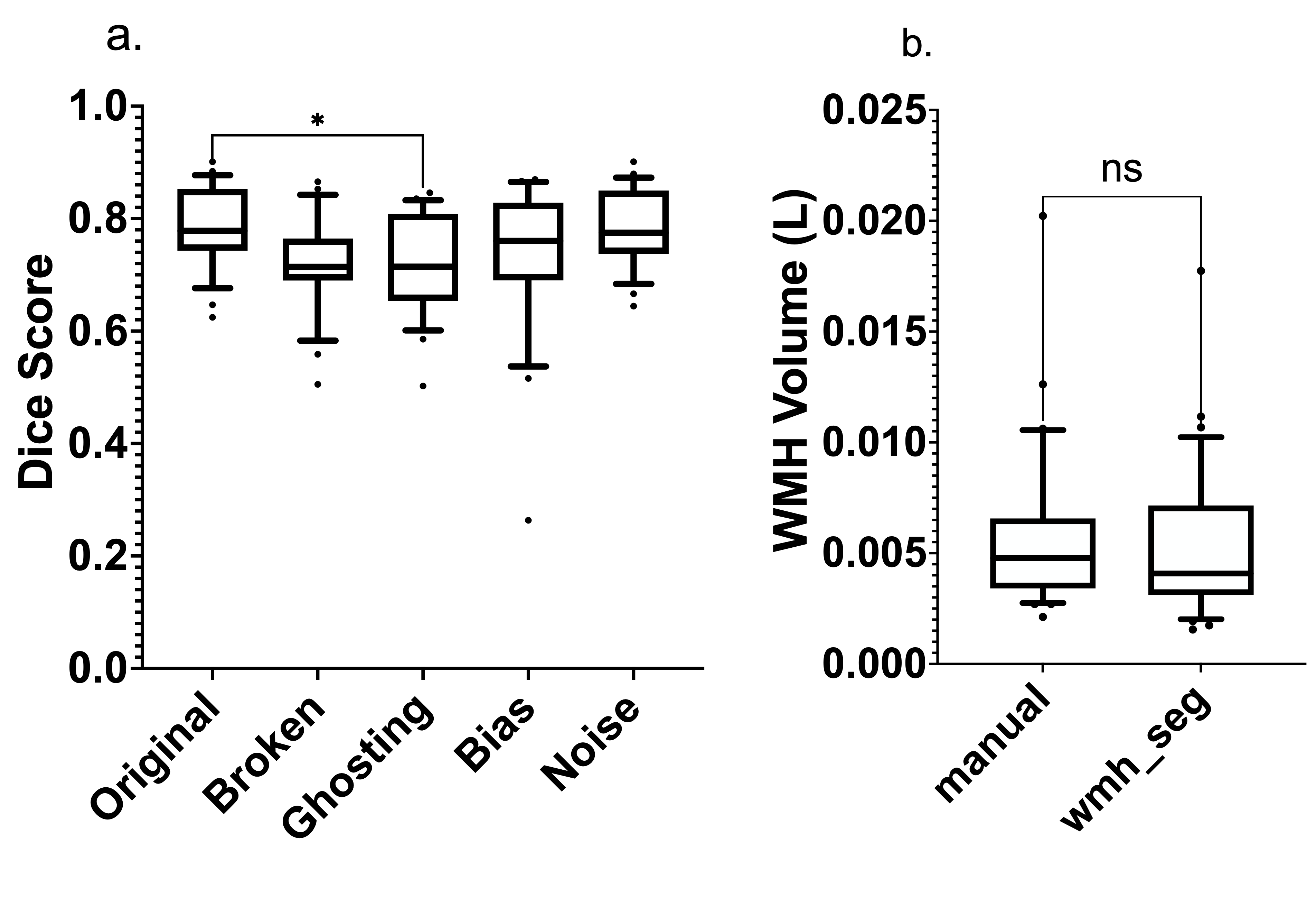}
\end{center}
   \caption{wmh\_seg quantitative white matter lesion segmentation on 7T FLAIR images and the images with the corresponding artifacts. a) wmh\_seg model inference on 7T FLAIR images alongside with corresponding artifacts. b) Volumetric comparison reveals no discernible differences between the manually labeled masks and the wmh\_seg segmentation outputs when applied to the uncorrupted 7T FLAIR images. }
\label{fig:fig8}
\end{figure}

\begin{figure}[!htb]
\begin{center}
   \includegraphics[width=1\linewidth]{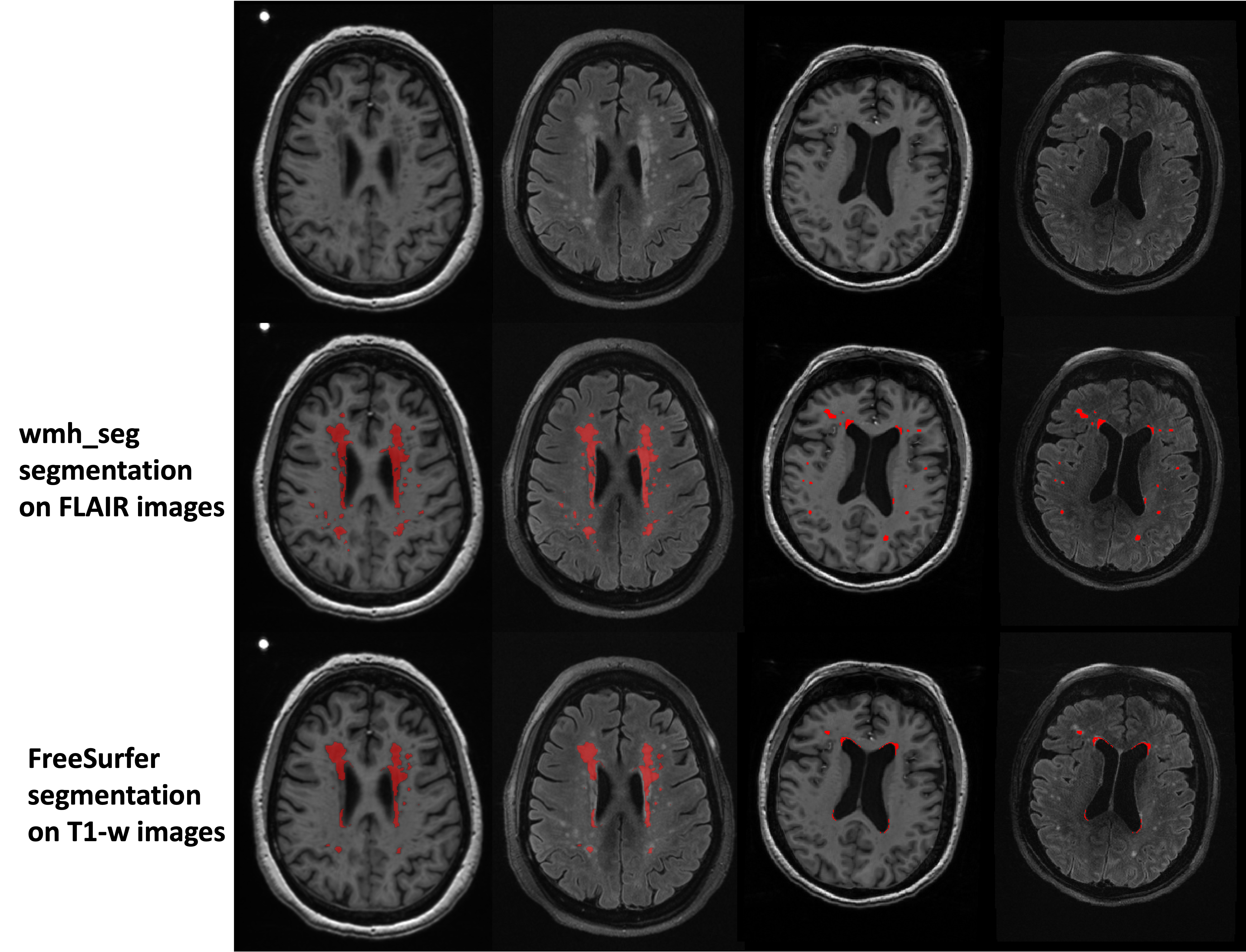}
\end{center}
   \caption{Qualitative white matter lesion segmentation comparison between FreeSurfer and our transformer-based model, overlayed on T1w and T2w FLAIR images respectively.}
\label{fig:fig9}
\end{figure}

\newpage
It's important to note that this doesn't undermine the effectiveness of the pgs model’s method; rather, it underscores the challenge of applying it to 7T images due to the distinctive image characteristics between 3T and 7T images. We would also like to highlight the sensitivity of the Dice score as a metric for quantifying segmentation outcomes. While our model may not attain the highest Dice coefficient for 7T images, it consistently demonstrates no volumetric disparities when compared to manually labeled masks. Lastly, in figure \ref{fig:fig9}, we compared our model’s output with the lesion segmentation on T1w images from FreeSurfer. 

Although, FreeSurfer is an excellent neuroimage segmentation tool, it only detects white matter lesions or white matter hypointensities on T1 weighted images which do not provide the best contrast between healthy tissue and morbid white matter lesions. As such FreeSurfer may fail to differentiate subcortical white matter lesions or hypointensities from the adjacent cortical gray matter on T1w images. As illustrated in Figure \ref{fig:fig9}, small, isolated foci of white matter lesions, usually in subcortical white matter or superficial white matter, were not identified using FreeSurfer.  

\newpage
\section{Discussion}
wmh\_seg aims to offer a robust WMH segmentation model that generalizes well across wide range of data with no preprocessing steps needed. The model was trained on T2-weighted FLAIR images at an unparalleled scale, including images acquired at 1.5 Tesla, 3 Tesla, and 7 Tesla ranging from 3 different scanner manufactures. Further, we augmented the training images by incorporating various MR artifacts. To date, this work is the first to offer white matter hyperintensity segmentation model that includes 7T FLAIR training images. During inference, wmh\_seg only requires the raw T2w FLAIR images. No preprocessing steps, hyperparameter inputs and other structural modalities are needed. This simple model inference guarantees standard quantification across users. Our results validated that wmh\_seg’s most robust generalizability across magnetic field strengths, scanner manufactures as well as MR image artifacts. Although, wmh\_seg shows accurate quantification on white matter hyperintensities, it was largely trained on normal aging participant images. White matter lesions induced by other disease pathologies such as stroke, and tumor may pose additional challenges for the model. Additional validation is needed on FLAIR images of other pathologies. We aim to release the trained model for public access, facilitating downstream fine-tuning for specific datasets.

\section{Conclusion}
In this study, we implemented a transformer-based deep learning model: wmh\_seg for white matter lesion segmentation on T2w FLAIR images that were acquired at diverse magnetic field strengths and institutes. Our additional data augmentation approach along with the diverse image sources improved the robustness of our model. Further, our segmentation results beat the WMH segmentation challenge winning team’s model inference on 7T images.  Lastly, qualitative inspection shows that our segmentation is more accurate in quantifying subcortical white matter lesions than FreeSurfer.  

\section{Author Contribution}
\textbf{Jinghang Li}: Methodology, Software, Formal analysis, Writing – original draft, Visualization, Data curation. \textbf{Tales Santini}: Methodology, Writing – review \& editing. \textbf{Yuanzhe Huang}: Methodology, Software. \textbf{Joseph M. Mettenburg}: Data curation, Writing – review \& editing, \textbf{Tamer S. Ibrahim}: Resources, Supervision. \textbf{Howard J. Aizenstein}: Resources, Supervision, Writing – review \& editing. \textbf{Minjie Wu}: Methodology, Resources, Project administration, Supervision, Writing – review \& editing.

\section{Acknowledgment}
This research was supported in part by the University of Pittsburgh Center for Research Computing: RRID:SCR\_022735, through the resources provided. Specifically, this work used the H2P cluster, which is supported by NSF award number OAC-2117681. This work was also supported by the National Institutes of Health under the grant number R01MH111265, R01AG067018 and RF1AG025516.


\bibliographystyle{unsrt}
\bibliography{bibliography.bib}

\begin{thebibliography}{10}

\bibitem{lee2016white}
Seonjoo Lee, Fawad Viqar, Molly~E Zimmerman, Atul Narkhede, Giuseppe Tosto, Tammie~LS Benzinger, Daniel~S Marcus, Anne~M Fagan, Alison Goate, Nick~C Fox, et~al.
\newblock White matter hyperintensities are a core feature of alzheimer's disease: evidence from the dominantly inherited alzheimer network.
\newblock {\em Annals of neurology}, 79(6):929--939, 2016.

\bibitem{polman2011diagnostic}
Chris~H Polman, Stephen~C Reingold, Brenda Banwell, Michel Clanet, Jeffrey~A Cohen, Massimo Filippi, Kazuo Fujihara, Eva Havrdova, Michael Hutchinson, Ludwig Kappos, et~al.
\newblock Diagnostic criteria for multiple sclerosis: 2010 revisions to the mcdonald criteria.
\newblock {\em Annals of neurology}, 69(2):292--302, 2011.

\bibitem{inzitari2009changes}
Domenico Inzitari, Giovanni Pracucci, Anna Poggesi, Giovanna Carlucci, Frederik Barkhof, Hugues Chabriat, Timo Erkinjuntti, Franz Fazekas, Jos{\'e}~M Ferro, Michael Hennerici, et~al.
\newblock Changes in white matter as determinant of global functional decline in older independent outpatients: three year follow-up of ladis (leukoaraiosis and disability) study cohort.
\newblock {\em Bmj}, 339, 2009.

\bibitem{lampe2019lesion}
Leonie Lampe, Shahrzad Kharabian-Masouleh, Jana Kynast, Katrin Arelin, Christopher~J Steele, Markus L{\"o}ffler, A~Veronica Witte, Matthias~L Schroeter, Arno Villringer, and Pierre-Louis Bazin.
\newblock Lesion location matters: the relationships between white matter hyperintensities on cognition in the healthy elderly.
\newblock {\em Journal of Cerebral Blood Flow \& Metabolism}, 39(1):36--43, 2019.

\bibitem{esiri1999cerebrovascular}
Margaret~M Esiri, Zsuzsanna Nagy, Maria~Z Smith, Lin Barnetson, and A~David Smith.
\newblock Cerebrovascular disease and threshold for dementia in the early stages of alzheimer's disease.
\newblock {\em The Lancet}, 354(9182):919--920, 1999.

\bibitem{heyman1998cerebral}
A~Heyman, GG~Fillenbaum, KA~Welsh-Bohmer, M~Gearing, SS~Mirra, RC~Mohs, BL~Peterson, and CF~Pieper.
\newblock Cerebral infarcts in patients with autopsy-proven alzheimer's disease: Cerad, part xviii.
\newblock {\em Neurology}, 51(1):159--162, 1998.

\bibitem{schneider2004cerebral}
JA~Schneider, RS~Wilson, JL~Bienias, DA~Evans, and DA~Bennett.
\newblock Cerebral infarctions and the likelihood of dementia from alzheimer disease pathology.
\newblock {\em Neurology}, 62(7):1148--1155, 2004.

\bibitem{snowdon1997brain}
David~A Snowdon, Lydia~H Greiner, James~A Mortimer, Kathryn~P Riley, Philip~A Greiner, and William~R Markesbery.
\newblock Brain infarction and the clinical expression of alzheimer disease: the nun study.
\newblock {\em Jama}, 277(10):813--817, 1997.

\bibitem{ben2011therapeutics}
Aliza~Bitton Ben-Zacharia.
\newblock Therapeutics for multiple sclerosis symptoms.
\newblock {\em Mount Sinai Journal of Medicine: A Journal of Translational and Personalized Medicine}, 78(2):176--191, 2011.

\bibitem{prins2004measuring}
ND~Prins, ECW Van~Straaten, EJ~Van~Dijk, M~Simoni, RA~Van~Schijndel, HA~Vrooman, PJ~Koudstaal, P~Scheltens, MMB Breteler, and F~Barkhof.
\newblock Measuring progression of cerebral white matter lesions on mri: visual rating and volumetrics.
\newblock {\em Neurology}, 62(9):1533--1539, 2004.

\bibitem{admiraal2005fully}
Faiza Admiraal-Behloul, DMJ Van Den~Heuvel, Hans Olofsen, Matthias~JP van Osch, Jeroen van~der Grond, Mark~A van Buchem, and Johan~HC Reiber.
\newblock Fully automatic segmentation of white matter hyperintensities in mr images of the elderly.
\newblock {\em Neuroimage}, 28(3):607--617, 2005.

\bibitem{iorio2013white}
Mariangela Iorio, Gianfranco Spalletta, Chiara Chiapponi, Giacomo Luccichenti, Claudia Cacciari, Maria~D Orfei, Carlo Caltagirone, and Fabrizio Piras.
\newblock White matter hyperintensities segmentation: a new semi-automated method.
\newblock {\em Frontiers in aging neuroscience}, 5:76, 2013.

\bibitem{schmidt2017lst}
Paul Schmidt and Lucie Wink.
\newblock Lst: A lesion segmentation tool for spm.
\newblock {\em Manual/Documentation for version}, 2:15, 2017.

\bibitem{wu2006fully}
Minjie Wu, Caterina Rosano, Meryl Butters, Ellen Whyte, Megan Nable, Ryan Crooks, Carolyn~C Meltzer, Charles~F Reynolds~III, and Howard~J Aizenstein.
\newblock A fully automated method for quantifying and localizing white matter hyperintensities on mr images.
\newblock {\em Psychiatry Research: Neuroimaging}, 148(2-3):133--142, 2006.

\bibitem{zwanenburg2010fluid}
Jaco~JM Zwanenburg, Jeroen Hendrikse, Fredy Visser, Taro Takahara, and Peter~R Luijten.
\newblock Fluid attenuated inversion recovery (flair) mri at 7.0 tesla: comparison with 1.5 and 3.0 tesla.
\newblock {\em European radiology}, 20:915--922, 2010.

\bibitem{kang2009hypertension}
Chang-Ki Kang, Chan-A Park, Hyon Lee, Sang-Hoon Kim, Cheol-Wan Park, Young-Bo Kim, and Zang-Hee Cho.
\newblock Hypertension correlates with lenticulostriate arteries visualized by 7t magnetic resonance angiography.
\newblock {\em Hypertension}, 54(5):1050--1056, 2009.

\bibitem{chen2023paired}
Xiaoyang Chen, Liangqiong Qu, Yifang Xie, Sahar Ahmad, and Pew-Thian Yap.
\newblock A paired dataset of t1-and t2-weighted mri at 3 tesla and 7 tesla.
\newblock {\em Scientific Data}, 10(1):489, 2023.

\bibitem{torrisi2018statistical}
Salvatore Torrisi, Gang Chen, Daniel Glen, Peter~A Bandettini, Chris~I Baker, Richard Reynolds, Jeffrey Yen-Ting Liu, Joseph Leshin, Nicholas Balderston, Christian Grillon, et~al.
\newblock Statistical power comparisons at 3t and 7t with a go/nogo task.
\newblock {\em NeuroImage}, 175:100--110, 2018.

\bibitem{ronneberger2015u}
Olaf Ronneberger, Philipp Fischer, and Thomas Brox.
\newblock U-net: Convolutional networks for biomedical image segmentation.
\newblock In {\em Medical Image Computing and Computer-Assisted Intervention--MICCAI 2015: 18th International Conference, Munich, Germany, October 5-9, 2015, Proceedings, Part III 18}, pages 234--241. Springer, 2015.

\bibitem{wu2019skip}
Jiong Wu, Yue Zhang, Kai Wang, and Xiaoying Tang.
\newblock Skip connection u-net for white matter hyperintensities segmentation from mri.
\newblock {\em IEEE Access}, 7:155194--155202, 2019.

\bibitem{kuijf2019standardized}
Hugo~J Kuijf, J~Matthijs Biesbroek, Jeroen De~Bresser, Rutger Heinen, Simon Andermatt, Mariana Bento, Matt Berseth, Mikhail Belyaev, M~Jorge Cardoso, Adria Casamitjana, et~al.
\newblock Standardized assessment of automatic segmentation of white matter hyperintensities and results of the wmh segmentation challenge.
\newblock {\em IEEE transactions on medical imaging}, 38(11):2556--2568, 2019.

\bibitem{xie2021segformer}
Enze Xie, Wenhai Wang, Zhiding Yu, Anima Anandkumar, Jose~M Alvarez, and Ping Luo.
\newblock Segformer: Simple and efficient design for semantic segmentation with transformers.
\newblock {\em Advances in Neural Information Processing Systems}, 34:12077--12090, 2021.

\bibitem{fischl2012freesurfer}
Bruce Fischl.
\newblock Freesurfer.
\newblock {\em Neuroimage}, 62(2):774--781, 2012.

\bibitem{griffanti2016bianca}
Ludovica Griffanti, Giovanna Zamboni, Aamira Khan, Linxin Li, Guendalina Bonifacio, Vaanathi Sundaresan, Ursula~G Schulz, Wilhelm Kuker, Marco Battaglini, Peter~M Rothwell, et~al.
\newblock Bianca (brain intensity abnormality classification algorithm): A new tool for automated segmentation of white matter hyperintensities.
\newblock {\em Neuroimage}, 141:191--205, 2016.

\bibitem{park2021white}
Gilsoon Park, Jinwoo Hong, Ben~A Duffy, Jong-Min Lee, and Hosung Kim.
\newblock White matter hyperintensities segmentation using the ensemble u-net with multi-scale highlighting foregrounds.
\newblock {\em Neuroimage}, 237:118140, 2021.

\bibitem{garcea2023data}
Fabio Garcea, Alessio Serra, Fabrizio Lamberti, and Lia Morra.
\newblock Data augmentation for medical imaging: A systematic literature review.
\newblock {\em Computers in Biology and Medicine}, 152:106391, 2023.

\bibitem{taylor2018improving}
Luke Taylor and Geoff Nitschke.
\newblock Improving deep learning with generic data augmentation.
\newblock In {\em 2018 IEEE symposium series on computational intelligence (SSCI)}, pages 1542--1547. IEEE, 2018.

\bibitem{perez2021torchio}
Fernando P{\'e}rez-Garc{\'\i}a, Rachel Sparks, and S{\'e}bastien Ourselin.
\newblock Torchio: a python library for efficient loading, preprocessing, augmentation and patch-based sampling of medical images in deep learning.
\newblock {\em Computer Methods and Programs in Biomedicine}, 208:106236, 2021.

\bibitem{raghu2021vision}
Maithra Raghu, Thomas Unterthiner, Simon Kornblith, Chiyuan Zhang, and Alexey Dosovitskiy.
\newblock Do vision transformers see like convolutional neural networks?
\newblock {\em Advances in Neural Information Processing Systems}, 34:12116--12128, 2021.

\bibitem{rajput4065778robustness}
Vishal Rajput.
\newblock Robustness of different loss functions and their impact on network's learning.
\newblock {\em Available at SSRN 4065778}.

\bibitem{kingma2014adam}
Diederik~P Kingma and Jimmy Ba.
\newblock Adam: A method for stochastic optimization.
\newblock {\em arXiv preprint arXiv:1412.6980}, 2014.

\end{thebibliography}

\end{document}